\documentclass{iaus}
\usepackage{graphicx}

\newcommand{\hi}{\mbox{H{\sc i}}}

\title[short title of paper]{\large Extended \hi\ Rotation Curve of M31 using deep DRAO observations}

\author[C. Carignan, L. Chemin \& T. Foster]   %% give here short author list %%
{\normalsize Claude Carignan$^1$, Laurent Chemin$^{1,2}$ \and Tyler Foster$^{3}$}
\affiliation{$^1$D\'ept. de Physique, Universit\'e de Montr\'eal, C.P. 6128, succ. centre-ville, 
Montr\'eal (Qc), Canada, H3C3J7; e-mail: carignan@astro.umontreal.ca\\[\affilskip]
$^2$G\'EPI, Observatoire de Paris, 5, pl. Janssen, 92195, Meudon, France; e-mail: laurent.chemin@obspm.fr \\[\affilskip]
$^3$Dept. of Physics \& Astronomy, Brandon University, Brandon, MB, Canada, R7A 6A9; e-mail: fostert@BrandonU.CA\\[\affilskip]}

\pubyear{2006}
\volume{235}  %% insert here IAU Symposium No.
\pagerange{1--1}
\jname{Galaxy Evolution across the Hubble Time}
\editors{F. Combes \& J. Palous, eds.}
\begin{document}

\maketitle
 \begin{abstract}
 Carignan et al. (2006) recently presented an extended \hi\ rotation curve (RC) of M31, using  
single dish observations from the 100m Effelsberg and Green Bank telescopes. These observations were 
motivated by a comparison with previous \hi\ data from Braun (1991) which presented a decreasing rotation 
curve as a function of radius. The Carignan et al. (2006) data were obtained along the semi-major axis of the approaching half of the M31 \hi\ disk and showed 
 a flat RC at large radius, extending up to $\sim$35 kpc (using $D=780$ kpc). 
  The kinematical analysis of M31 is pursued here and  new deep 21cm observations 
  obtained at the Dominion Radio Astrophysical Observatory (DRAO) are presented. 
   A tilted-ring model is fitted to a new \hi\ velocity field, allowing the derivation of 
   the position angle $P.A.$ and inclination $i$ as a function of radius.  
 We concentrate on the approaching half of the disk in order to compare our new results with those from Carignan et al (2006). 
It is shown that the disk warping of M31 does not severely contaminate the kinematics of the neutral gas. 
  As a consequence, the RC from the single dish data is in very good agreement with the newly derived RC. 
\end{abstract}
 
 \section{Observations, results and future work}
\begin{figure}[!h]\centering  
\includegraphics[width=\columnwidth]{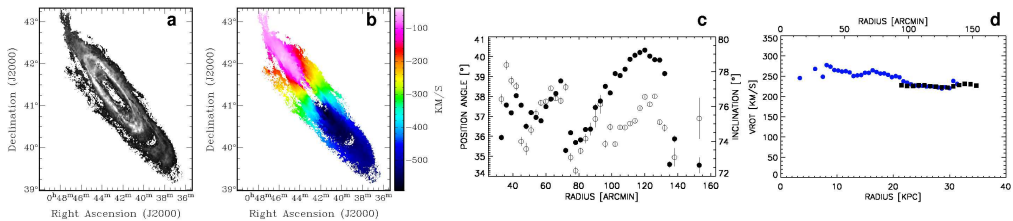}
\caption{\textbf{a-b} \hi\ integrated intensity map and velocity field of M31 from new DRAO observations. 
\textbf{c} Radial profiles of $P.A.$ and $i$  (full and empty dots resp.). \textbf{d} 
New \hi\ rotation curve of M31 (circles) and comparison with recent single dish velocities (squares).}
\label{fig1}
\end{figure}

In order to recover the flux from all the angular scales and a complete velocity field, 
21cm data were obtained using the DRAO synthesis telescope (Landecker et al. 2000), combined with the DRAO 26m single dish. 
Details of the data reduction procedure can be found in Taylor et al. (2003).
Gaussian profiles were fitted to the \hi\ data-cube to extract the different moment maps.
Figures~\ref{fig1}a and \ref{fig1}b present the provisional integrated intensity map of the 21cm emission line
and its corresponding velocity field. 
Because M31 is known to have a warped disk, a tilted ring model has to 
be fitted to the velocity field. It allows to derive accurate kinematical parameters and rotation velocities 
as a function of radius. We focus on the approaching half of the disk in order to compare with the RC from Carignan et al. (2006).
Figure~\ref{fig1}c shows that $i$ does not vary significantly as a function of radius, whereas 
the $P.A.$ variation is more important. In other words, the warp of M31 does not contaminate 
so much the \hi\ disk kinematics at large radius, at least in its approaching half.  
As a consequence, the RC from Carignan et al. (2006) is very similar to our new measurement (Fig~\ref{fig1}d). 
Future work will compare the receding and approaching sides' kinematics in order to derive 
a global RC and a mass model for M31 (Chemin et al. 2007, in prep.).

\end{document}